\documentclass[a4paper,11pt]{article}
\pdfoutput=1 

\usepackage{jcappub} 

\usepackage[T1]{fontenc} 

\usepackage{amsmath, amssymb, amsthm, graphicx, epsfig, fancyhdr,epsfig,multirow}
\usepackage[utf8]{inputenc}
\usepackage{amsmath}
\usepackage{amsfonts}
\usepackage{amssymb}
\usepackage{xcolor}
\usepackage{comment}
\usepackage{subcaption}
\usepackage[normalem]{ulem}
\usepackage[left=2cm,right=2cm,top=2cm,bottom=2cm]{geometry}
\usepackage[font=small,labelfont=bf,textfont=it]{caption}
\usepackage[toc,page]{appendix}
\usepackage{feynmf}
\usepackage{commath}
\usepackage{tikz}
\usepackage{hyperref}
\usepackage{slashed}
\usetikzlibrary{decorations.pathmorphing,decorations.markings,plotmarks,quotes,angles}

\newcommand{\bea}{\begin{eqnarray}}
\newcommand{\eea}{\end{eqnarray}}

\title{\boldmath 
A congruous approach with realistic cross section towards limiting sub-GeV dark matter from LUX-ZEPLIN
}

\author[a]{Atanu Guha}
\author[a]{and Jong-Chul Park}

\affiliation[a]{Department of Physics and Institute of Quantum Systems, \protect\\ Chungnam National University, Daejeon, 34134, Republic of Korea}

\emailAdd{atanu@cnu.ac.kr}
\emailAdd{jcpark@cnu.ac.kr}

\abstract{We present constraints on sub-GeV dark matter (DM) through the mechanism of being boosted by cosmic rays (CRs). We utilize the nuclear recoil data from the LUX-ZEPLIN (LZ) experiment for this purpose. Without the mechanism of boosted dark matter (BDM), sub-GeV DM particles are cold enough to produce detectable nuclear recoil in the LZ experiment above the detector threshold. We choose to work with the leading components of cosmic rays to take into account the boost due to them towards the cold DM. In the present discussion we worked on models consisting of a Dirac fermion $\chi$ with a new $U(1)'$ gauge symmetry and DM particles have non-zero coupling to the nucleons as per the model parameters. Specific examples of the energy dependence of the scattering cross section have been invoked through the secluded dark photon model and $U(1)_{B-L}$ model. Additionally, we present the upper bound on the interaction cross section due to the Earth shielding effect in the light of a systematic analysis of the energy loss by the BDM while traveling to the underground detector through the Earth's crust.}

\begin{document}
\maketitle
\flushbottom

\section{Introduction}

 The presence of a substantial amount of dark matter (DM) in our Universe and dominance over the baryonic matter is deep-rooted over the last few decades~\cite{Bertone:2004pz,Lisanti:2016jxe,Bauer:2017qwy,Cirelli:2024ssz}. Strong indications of the basis of such believes come naturally from numerous cosmological and astrophysical observations which include rotation curves of the galaxies~\cite{Rubin:1970zza,Rubin:1978kmz,Freeman:1970mx}, gravitational lensing effects~\cite{Brainerd:1995da,DellAntonio:1996ylb,Hoekstra:2003pn}, cosmic microwave background radiation~\cite{Planck:2015fie}, analysis of the X-ray and lensing observations of the bullet cluster~\cite{Clowe:2006eq,Randall:2008ppe} etc. At large scales, the evidences are quite unambiguous and elucidated by gravitational forces despite the particle nature of DM candidates and their interaction strengths are yet to be emerged. There have been tireless efforts in this direction, both theoretically and experimentally, through various avenues. Some leading experiments, namely indirect searches, look for the standard model (SM) remnants of the annihilating DM particles~\cite{Fermi-LAT:2009ihh,AMS:2014bun,AMS:2014gdf,Boudaud:2016mos,IceCube:2014stg,DAMPE:2017fbg,PAMELA:2013vxg,PAMELA:2008gwm}. On the other hand, following~\cite{Goodman:1984dc} the recoil spectrum of the SM target particles are being observed in the direct detection (DD) experiments while the DM particles scatter off them~\cite{DAMA:2008jlt, CRESST:2015txj, DarkSide:2018ppu, SuperCDMS:2018mne, XENONCollaboration:2023orw, LUX:2022vee, PandaX-II:2021kai}. 
 
 Till date the strongest bound on the spin-independent elastic scattering cross section of cold DM with nucleon is reported as $ \approx 5 \times 10^{-48}~\rm{cm^2}$ for DM mass of 30 GeV~\cite{LZ:2022lsv}, whereas the most stringent constraint on the DM-electron elastic scattering cross section is known to be $ \approx 5 \times 10^{-41}~\rm{cm^2}$ at 200 MeV of the DM mass~\cite{XENON:2022ltv}. But for low mass DM the search strategy endures drawback. Conventional limits on the DM-SM cross section from DD searches rely on the cold DM consideration where the average velocity of halo DM is approximately $10^{-3}c$, $c$ being the speed of light. Due to this, a drastic loss of sensitivity is observed at DD experiments for low mass DM. Light DM particles suffer from the phenomena of less efficient energy transfer which in turn induces recoil energy of the target particles below the detector threshold. Considering DM-nucleon spin-independent interaction the lowest explored mass by the traditional DD experiments is approximately 0.3 GeV~\cite{CRESST:2019jnq} whereas in the case of DM-electron interaction the same is around 0.3 MeV~\cite{EDELWEISS:2020fxc,SuperCDMS:2020ymb}. In our present study we mainly focus on the LZ experiment which quantitatively loses sensitivity for sub-GeV DM in case of nuclear recoils and for sub-MeV DM in case of electron recoils.
 
 In order to explore the low mass region of the parameter space, several alternatives have been developed. The idea of boosted dark matter (BDM) has been quite popular and seemed feasible since its conceptual introduction in recent literature. BDM has been introduced through different classes of theoretical modeling, e.g., the two component DM models with a hierarchical mass spectrum~\cite{Giudice:2017zke,Chatterjee:2018mej,Kim:2018veo,Kim:2019had} and cosmic ray boosted DM (CRBDM)~\cite{Ema:2018bih,Cho:2020mnc,Bringmann:2018cvk,Xia:2021vbz,Cappiello:2019qsw,Bardhan:2022bdg,Ghosh:2021vkt,PandaX-II:2021kai,Cao:2020bwd,Guha:2024mjr,Bell:2023sdq,Guo:2020oum,Cappiello:2024acu, Jho:2021rmn, Jho:2020sku, Das:2021lcr, Maity:2022exk, Alvey:2022pad, Diurba:2025lky, Alhazmi:2025nvt, Kumar:2024xfb, Das:2024ghw}. In this work we investigate the effect of CRBDM on the nuclear recoil spectrum of the LZ experiment while taking into account the leading components of cosmic rays~\cite{Workman:2022ynf,PhysRevD.98.030001}. In~\cite{Guha:2024mjr} we already showed that considering the boosting mechanism and working in a theoretical framework to invoke energy dependence into the interaction cross section, we could achieve significant improvements on the exclusion limits compared to the energy independent constant cross section scenario. While without diving into much details of any specific underlying model, the constant cross section consideration is capable of providing a quick but reasonable estimation about the exclusion limit of the DM-SM interaction cross section for a specific DM mass, but for the study of a realistic scenario a proper theoretical modeling is required. We estimate the exclusion limits on DM-nucleon interaction cross section working with the models having a a new $U(1)'$ gauge symmetry and the theory contains a new Dirac fermion $\chi$. We choose two benchmark model for this purpose, namely, secluded dark photon model and $U(1)_{B-L}$ model. The DM fermion $\chi$ interacts with the SM fermions through the new gauge boson mediator $A'$. In particular $A'$ is kinetically mixed with the SM $U(1)_Y$ gauge boson within the framework of secluded dark photon model (the kinetic portal) whereas, the conservation of the difference between the baryon number and the lepton number is encoded into the new $U(1)'$ gauge symmetry in case of the $U(1)_{B-L}$ model. DM fermions are charged under the new $U(1)'$ for both the models. The strength of the DM-nucleus interactions depends upon the total electric charge or the total baryon number of the nucleus, within the framework of the secluded dark photon model and $U(1)_{B-L}$ model, respectively. The two models stated above, are capable of depicting DM-SM interactions through the new $U(1)'$ gauge boson as the mediator. We found that the energy dependence of the cross section plays a crucial role in improving the constraints.
 
 We organize the subsequent discussion as follows. In Sec.~\ref{sec:CR_composition_flux} we briefly summarize the leading components of cosmic ray and estimate the corresponding fluxes. We describe the benchmark models we choose to present in this work in Sec.~\ref{sec:benchmark}. In Sec.~\ref{sec:BDM_flux} we estimate the flux of cosmic ray boosted DM (CRBDM) from the knowledge of the fluxes of the different components of cosmic ray. We write down the expression for the calculation of event rate in Sec.~\ref{sec:event_rate}. In Sec.~\ref{sec:constraint_eps} we constrain the dark sector coupling through two different approaches using LZ data. We translate the above-mentioned bounds to the DM-nucleon interaction cross section vs DM mass plane for the corresponding models in Sec.~\ref{sec:cs_constraint}. Finally, we conclude in Sec.~\ref{sec:conclusion}.

\section{Cosmic Ray : Composition and Fluxes}
\label{sec:CR_composition_flux}

 By definition and detection strategies, charged particles constitute cosmic rays. Earth-based detectors infer about the composition of the cosmic rays and the fluxes of the individual components. Dominant contributions to the cosmic rays come from protons, electrons, and helium nuclei~\cite{Longair:1992ze}. In our present study, we consider the major contributions only from these leading components to boost the cold DM. Details of the composition have already been discussed to some extent in~\cite{Guha:2024mjr,Workman:2022ynf}.

 Following~\cite{PhysRevD.98.030001,Boschini:2018zdv} we obtain the fluxes of the cosmic rays component-wise. In our calculation, we deploy the parameterization of the leading components of CR, viz., electron, proton, and Helium nuclei as described below. The local interstellar spectrum of the CR electron has been parameterized by~\cite{Boschini:2018zdv} which gives the electron flux in a simplified manner, as shown in Eq.~(~\ref{Eq:CRe_flux_parameterization}).
\begin{align}
  \frac{d\Phi_e}{dT_e}(T_e) = \begin{cases}
     \mbox{\Large\( \frac{1.799 \times 10^{44}~ T_e^{-12.061}}{1+ 2.762 \times 10^{36}~ T_e^{-9.269} + 3.853 \times 10^{40}~ T_e^{-10.697}} \)} & \text{if $T_e < 6880$ MeV} \\
   \mbox{\tiny\( \)} & \mbox{\tiny\( \)} \\
      3.259 \times 10^{10}~T_e^{-3.505} + 3.204 \times 10^{5}~T_e^{-2.620} & \text{if $T_e \geqslant 6880$ MeV}
    \end{cases}
\label{Eq:CRe_flux_parameterization}
\end{align}
where the unit of $\frac{d\Phi_e}{dT_e}(T_e)$ is given in $\rm{\left(m^2~s~sr~MeV \right)^{-1}}$ and the kinetic energy ($T_e$) of the CR electrons is in MeV. This parameterization agrees well with the observational data from Fermi-LAT~\cite{Fermi-LAT:2011baq,Fermi-LAT:2009yfs,Fermi-LAT:2010fit,Fermi-LAT:2017bpc}, AMS-02~\cite{AMS:2014gdf}, PAMELA~\cite{PAMELA:2011bbe,CALET:2017uxd}, and Voyager~\cite{Cummings:2016pdr,Stone:2013}.

On the other hand, we estimate the differential intensity ($\frac{dI}{dR}$) in terms of rigidity ($R$) for CR proton and Helium nuclei following~\cite{Boschini:2017fxq,DellaTorre:2016jjf}. 

 \begin{align}
  \frac{dI}{dR} \times R^{2.7} = \begin{cases}
     \sum_{i=0}^{5} a_i R^i  , & \text{if $R \leqslant 1$ GV} \\
   \mbox{\tiny\( \)}  & \mbox{\tiny\( \)} \\
      \mbox{\Large\( b + \frac{c}{R} + \frac{d_1}{d_2 + R} + \frac{e_1}{e_2 + R} + \frac{f_1}{f_2 + R} + gR \)}, & \text{if $R > 1$ GV}
    \end{cases},
\label{Eq:CRp_flux_parameterization}
\end{align}
where, the values of the parameters $a_i, b, c, d_1, d_2, e_1, e_2, f_1, f_2, g$ are given in~\cite{Boschini:2017fxq,DellaTorre:2016jjf,Guha:2024mjr}.
  
Then we plug-in the above parameterization for $\frac{dI}{dR}$ into the following expressions and estimate CR proton and Helium flux.
  \bea
  \frac{d\Phi_p}{dT_p}(T_p) = 4 \pi \frac{dR}{dT_p} \frac{dI}{dR} \nonumber \\
  \frac{d\Phi_{He}}{dT_{He}}(T_{He}) = 4 \pi \frac{dR}{dT_{He}} \frac{dI}{dR}
  \eea

In our present work, DM is boosted by the CR electrons and nucleons in the framework of the models described in Sec.~\ref{sec:benchmark}. From the estimation of the CR fluxes we evaluate the BDM flux in Sec.~\ref{sec:BDM_flux}.

\section{Benchmark models and the state-of-the-art constraints} 
\label{sec:benchmark}

In this section, we provide the details of the benchmark models we work on. After introducing them in an unambiguous manner, we also discuss the constraints on the model parameters present in the recent literature. The focus of this extensive analysis revolves around the models consisting of additional $U(1)'$ symmetry. We investigate the CRBDM flux and thereafter the exclusion limit on the energy dependent cross section. We invoke the energy dependence into the interaction cross section working within the framework of the secluded dark photon and $U(1)_{B-L}$ model. In light of the proper formulation of the underlying models, the final results are discussed in Sec.~\ref{sec:result}.

\subsection{Models with new $U(1)'$ gauge symmetry}  
\label{sec:U(1)_models}

We work on the following two popular models. Both the models consists of dark fermions and a new vector boson as mediator.
 
 \begin{itemize}
 
 \item Secluded dark sector : This model consists of a spontaneously broken new $U(1)'$ gauge symmetry which has a kinetic mixing with the SM $U(1)_Y$ symmetry.
\bea
 \mathcal{L} &=& \mathcal{L}_{SM} + \bar{\chi} \left(i \slashed{\partial} - m_\chi \right) \chi - g_{\chi} \bar{\chi} \gamma_\mu \chi \hat{A'}^\mu \nonumber \\ &+& \frac{1}{2} m_{\hat{A'}}^2 \hat{A'}_\mu \hat{A'}^\mu  - \frac{1}{4} \hat{A'}_{\mu \nu} \hat{A'}^{\mu \nu} - \frac{\sin \varepsilon}{2} \hat{B}_{\mu \nu} \hat{A'}^{\mu \nu}
 \eea 
 
 where, $A'$ is the new gauge boson and $\chi$ is the DM fermion. The mass $m_{\hat{A'}}$ is generated due to the spontaneous symmetry breaking. The $U(1)_Y \times U(1)'$ kinetic mixing produces a non-zero coupling between the dark photon ($A'$) and the standard model fermions~\cite{Chun:2010ve,Fabbrichesi:2020wbt,Gu:2017gle}. By redefinition of the fields, we diagonalize away the kinetic mixing term following~\cite{Chun:2010ve} and write the relevant interaction terms
 \bea
 \mathcal{L} &\supset& {A'}_\mu \left[ g_{fL}^{A'}~\bar{f} \gamma^\mu P_L f + g_{fR}^{A'}~\bar{f} \gamma^\mu P_R f + g_{\chi}^{A'}~\bar{\chi} \gamma^\mu \chi \right] \nonumber \\
 &+& {Z}_\mu \left[ g_{fL}^{Z}~\bar{f} \gamma^\mu P_L f + g_{fR}^{Z}~\bar{f} \gamma^\mu P_R f + g_{\chi}^{Z}~\bar{\chi} \gamma^\mu \chi \right]
 \eea 
 where, $f$ is the standard model fermions and the corresponding couplings are taken from~\cite{Chun:2010ve}. The DM fermions ($\chi$) couple to the charged SM fermions through the hypercharge quantum number with $A^{\prime}$ as the mediator. As a result in case of DM-nuclear interaction the cross section is related to the DM-nucleon cross section by a factor which is proportional to the square of the number of protons inside the corresponding nucleus.

 \item $U(1)_{B-L}$  : 
 In this model the new $U(1)'$ gauge symmetry stands for the conservation of the difference between the baryon number and the lepton number. Both the SM fermions and the DM fermions are charged under the new $U(1)'\equiv U(1)_{B-L}$ symmetry~\cite{Bauer:2018onh}.  
 
\bea
\mathcal{L}_{B-L} \supset g_{B-L} \left[- \bar{l} \gamma^\mu A^{'}_{\mu} l - \bar{\nu}_R \gamma^\mu A^{'}_{\mu} \nu_R + \frac{1}{3} \bar{q} \gamma^\mu A^{'}_{\mu} q \right] \nonumber \\
 - g_{\chi} \bar{\chi} \gamma_\mu \chi {A'}^\mu
\eea

All SM the leptons and baryons are coupled to the DM fermions ($\chi$) through the new mediator ($A^{\prime}$) via their baryonic or leptonic quantum number. So in case of the DM-nuclear interaction the cross section is enhanced from the DM-nucleon cross section by a factor which is proportional to the square of the total number of baryons inside the corresponding nucleus.

\end{itemize}

The two models stated above, are capable of depicting DM-SM interactions through the new $U(1)'$ gauge boson as the mediator. We use the above mentioned models to estimate the BDM flux (Sec.~\ref{sec:BDM_flux}) and predicted event rates (Sec.~\ref{sec:event_rate}) at LZ experiment in the following discussions.

\subsection{Existing constraints on the model parameters from previous studies and experiments}  
\label{sec:existing_constraints}

 In recent times, there have been attempts to constrain the model parameters both theoretically and experimentally. We summarize such aspects in the light of the above two benchmark models and present them along with our estimated constraints using LZ data. Since we consider the new vector boson in our framework to be the mediator between the DM and SM fermions, the overall coupling between them is constrained from various experiments for different values of the mass of the vector boson. The estimation of the experimental bound depends on the consideration of the specific scenario of whether the dominant decay channel of the vector boson is visible or invisible. Depending on whether the new boson decays dominantly to the dark sector particles (invisibly) or to the standard model particles (leptons and quarks), there exist exclusion limits of two categories, viz. invisible decay mode and visible decay mode~\cite{Inan:2021dir,Filippi:2020kii,Gninenko:2020hbd,Bauer:2018onh}. A concise description of the same has been presented in~\cite{Guha:2024mjr}.
 
 To show our estimated bounds on the coupling parameters between DM and SM particles for the benchmark models, along with the already constrained region of the parameter space, we combine the strongest existing constraints obtained from different beam-dump, collider, and the Super Proton Synchrotron (SPS) experiments into the consolidated version in Fig.~\ref{fig:coup_constraint_mvecA}. The exclusion limits in the scenario of invisible decay mode, have been collected from the SPS experiments (NA64~\cite{Andreev:2021fzd,Banerjee:2019pds}, NA62~\cite{NA62:2019meo}), collider experiment (BaBar~\cite{Filippi:2019lfq,BaBar:2017tiz}), and electron beam-dump experiments (E787 and E949~\cite{Essig:2013vha}). On the other hand, we have gathered and combined the strongest exclusion bounds for the visible decay mode from the SPS experiments (NA64~\cite{NA64:2019auh}, NA48~\cite{NA482:2015wmo}, WASA~\cite{WASA-at-COSY:2013zom}), collider experiments (BESIII~\cite{BESIII:2017fwv}, Babar~\cite{Filippi:2019lfq,BaBar:2014zli}, LSND~\cite{LSND:1997vqj}, HADES~\cite{HADES:2013nab}, KLOE~\cite{KLOE-2:2012lii,KLOE-2:2014qxg,Anastasi:2015qla}), electron beam-dump experiments (E141~\cite{Fabbrichesi:2020wbt,Riordan:1987aw}, E137~\cite{Fabbrichesi:2020wbt,Bjorken:1988as,Batell:2014mga,Marsicano:2018krp}, E774~\cite{Bross:1989mp}), proton beam-dump experiments(nu-Cal~\cite{Blumlein:2011mv,Blumlein:2013cua}, CHARM~\cite{Gninenko:2012eq}), and electron-nucleus fixed target scattering experiments (APEX~\cite{APEX:2011dww}, A1~\cite{Merkel:2014avp}). 
For more details, please refer to Refs.~\cite{Inan:2021dir,Filippi:2020kii,Gninenko:2020hbd,Bauer:2018onh} and Section~\ref{sec:constraint_eps}.

In addition to the existing limits described above, we also incorporate some of the astrophysical constraints. In particular, for stellar cooling and supernova cooling scenario, the contribution of the new physics channel to the cooling mechanism through the new vector boson is constrained in terms of the maximum allowed energy loss rate for the new channel. As a consequence, the couplings are constrained in order to maintain the upper limit of the energy loss rate~\cite{Chang:2016ntp, Rrapaj:2015wgs, Hardy:2016kme}. The energy loss rate criteria for different class of stars and supernova are given as the following description~\cite{Hardy:2016kme}, where, $\epsilon_{new}$ is the energy loss rate through the new physics channel.

\begin{itemize}
    \item For the Sun : $\epsilon_{new} \lesssim 0.2~\rm{erg~gm^{-1}~s^{-1}}$.
    \item For Red Giant (RG) stars : $\epsilon_{new} \lesssim 10~\rm{erg~gm^{-1}~s^{-1}}$
    \item For Horizontal Branch (HB) stars : $\epsilon_{new} \lesssim 10~\rm{erg~gm^{-1}~s^{-1}}$
    \item For the core-collapse Supernova like SN1987A : $\epsilon_{new} \lesssim 10^{19}~\rm{erg~gm^{-1}~s^{-1}}$
\end{itemize}

\section{CRBDM Formalism}
\label{sec:BDM_flux}

 In this section, we describe the estimation of the flux of the CRBDM in detail. As a first step, we use the flux of the CRs and evaluate the CRBDM flux. Then we also take into consideration the shielding effect at the detector as a result of the traveling of CRBDM through the Earth's crust. Due to this phenomenon, the CRBDM flux gets attenuated before reaching the detector.

\subsection{Flux of CRBDM due to the upscattering by CRs}

 The interactions described in Sec.\ref{sec:benchmark} equip the CRs to produce non-zero population of the boosted DM. We take into consideration the dominant ingredients of CRs as the major boosting agents for galactic cold DM and we estimate the flux of CRBDM using the following expression.    
  \bea
 \frac{d\Phi_\chi}{dT_\chi} &=& D_{eff} \times \frac{\rho_\chi^{\rm{local}}}{m_\chi}~  \Biggl[ \int^{\infty}_{T_e^{min}} dT_e ~ \frac{d\Phi_e}{dT_e}~ \frac{d \sigma_{\chi e}}{dT_\chi} \nonumber \\ &+& \int^{\infty}_{T_p^{min}} dT_p ~ \frac{d\Phi_p}{dT_p}~ \frac{d \sigma_{\chi p}}{dT_\chi} G_{p}^2(2m_\chi T_\chi) \nonumber \\ &+& \int^{\infty}_{T_{He}^{min}} dT_{He} ~ \frac{d\Phi_{He}}{dT_{He}}~ \frac{d \sigma_{\chi {He}}}{dT_\chi} G_{He}^2(2m_\chi T_\chi) \Biggr] \nonumber \\
 \label{Eq:DMflux-wrt-T-full}
 \eea
 with 
 \bea
 \frac{d \sigma_{\chi i}}{dT_\chi} = {g_{i}^{A'}}^2 {g_{\chi}^{A'}}^2 \frac{2m_\chi \left(m_i + T_i \right)^2 - T_\chi \left\lbrace \left( m_i + m_\chi \right)^2 + 2 m_\chi T_i  \right\rbrace + m_\chi T_\chi^2}{4 \pi \left(2 m_i T_i + T_i^2 \right) \left(2 m_\chi T_\chi + m_{A'}^2 \right)^2}; ~~i=e,p,He \nonumber \\
 \label{eq:diff_cs_p}
 \eea
 
 and $G_i$ be the nucleon electromagnetic form factors. Sec.~\ref{sec:CR_composition_flux} facilitates us with the flux of CR electrons,protons and Helium nuclei. $D_{eff}$ is the effective diffusion zone parameter (in this work taken as 10 kpc) and $\rho_\chi^{\rm{local}} = 0.3~\rm{GeV.cm^{-3}}$ is the local mass density of the cold DM halo. $G_i$ for the hadronic elastic scattering has been provided in~\cite{Bringmann:2018cvk,Perdrisat:2006hj,Lei:2020mii}
 \bea
 G_i\left(q^2\right) = \left(1+ \frac{q^2}{\Lambda_i^2} \right)^{-2}
 \eea
 where, $\Lambda_p= 770~\rm{MeV}$, $\Lambda_{He}= 410~\rm{MeV}$ and $q$ be the momentum transfer. The differential scattering cross section in the point-like limit is then gives as 
 \bea
 \frac{d\sigma_{\chi i}}{d\Omega} = \frac{d\sigma_{\chi i}}{d\Omega}\Bigg|_{q^2=0} G_i\left(2m_\chi T_\chi\right)
 \eea
The energy transfer by a CR particle $i$ to galactic cold DM is estimated by kinematic analysis as~\cite{Ema:2018bih,Bringmann:2018cvk,Cappiello:2019qsw,Guha:2024mjr}
    \bea
  T_{\chi} &=&  T_{\chi}^{max}  \left( \frac{ 1 - \cos \theta }{2}\right)\nonumber \\ T_{\chi}^{max} &=& \frac{\left(T_i\right)^2+ 2 T_i m_i  }{T_i+ \left(m_i +m_{\chi} \right)^2/\left(2 m_{\chi} \right)} 
  \label{Eq:T-chi-max} 
  \eea
 where $\theta$ is the scattering angle at the center of the momentum frame. In order to get the values of $T_i^{min}$ (used in Eq.(\ref{Eq:DMflux-wrt-T-full})) we need to solve Eq.(\ref{Eq:T-chi-max}) for a fixed $T_\chi$ 
  \bea
  T_i^{min} = \left( \frac{T_{\chi}}{2} -m_i \right) \left[ 1 \pm \sqrt{ 1 +  \frac{2 T_{\chi}}{ m_{\chi}} \frac{\left(m_i +m_{\chi}\right)^2}{\left( 2m_i - T_{\chi} \right)^2}} \right] 
  \label{Eq:T-nu-min}
  \eea
  The $+$ and $-$ sign in Eq.~(\ref{Eq:T-nu-min}) stand for the scenario where $T_\chi > 2 m_i$ and $T_\chi < 2 m_i$, respectively. 
 
  For two different benchmark models, the corresponding coupling schemes used in Eq.~(\ref{eq:diff_cs_p}) are defined as 
 
 \begin{itemize}
 \item Secluded $U(1)'$ : 
$ \abs{g_{e}^{A'}} = \abs{g_{p}^{A'}} =  e \varepsilon \cos \theta_W $ , $\abs{g_{He}^{A'}} = 2 \abs{g_{p}^{A'}}$, $g_{\chi}^{A'} = g_{\chi}$;  $\theta_W $ be the Weinberg angle.
 \item $U(1)_{B-L}$ : $ \abs{g_{e}^{A'}} = \abs{g_{p}^{A'}} =  g_{B-L} $ , $\abs{g_{He}^{A'}} = 4 \abs{g_{p}^{A'}}$, $g_{\chi}^{A'} = g_{\chi}$
 \end{itemize}

\begin{figure*}
 \centering
 \includegraphics[width=0.49 \textwidth]{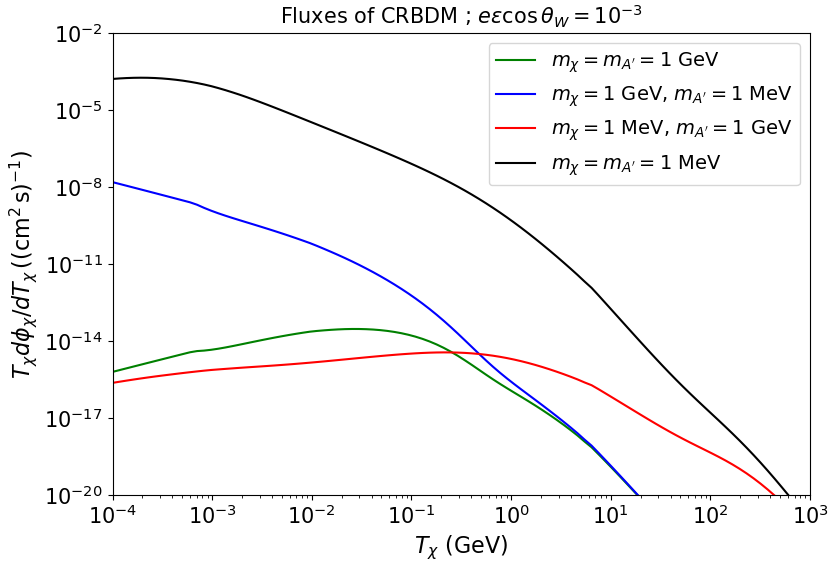}
 \includegraphics[width=0.49 \textwidth]{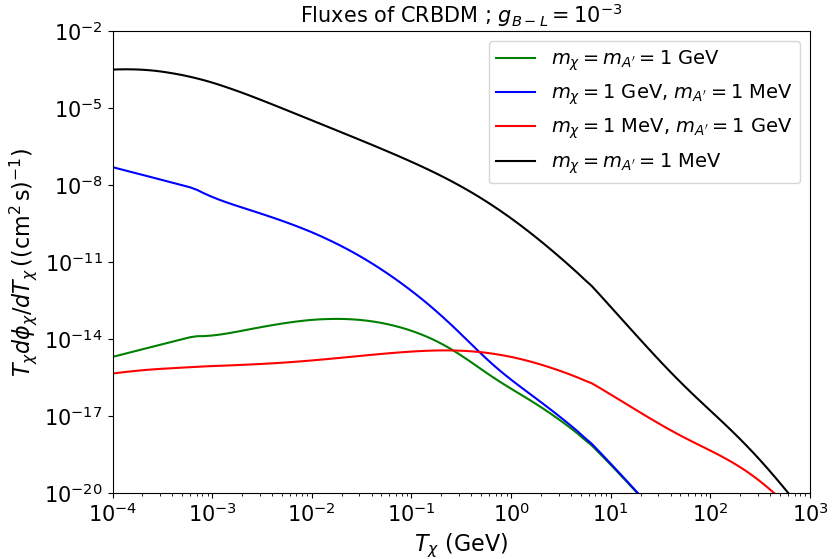}
 \caption{Cosmic ray electron, proton and Helium nuclei boosted DM flux for various DM mass and mediator mass considering $g_\chi = 1,  e \varepsilon \cos \theta_W = 10^{-3}$ (left panel) and $g_\chi = 1, g_{B-L} = 10^{-3}$ (right panel).
 }
 \label{fig:flux_U1BL_secluded}
 \end{figure*}

 We present a visual comparison of the BDM fluxes corresponding to various combinations of the DM mass and mediator mass for a specific parameter set in Fig.~\ref{fig:flux_U1BL_secluded}.
 
To ensure the CR proton elastic scattering with cold DM and to avoid the deep inelastic scattering, we restrict the maximum momentum transfer to be $q_p < 2$ GeV. 

\bea
q_p < 2~\rm{GeV} \nonumber \\
\implies \sqrt{2 m_\chi T_\chi} < 2~\rm{GeV} \nonumber \\
\implies T_\chi^{max} = \frac{\left(2~\rm{GeV} \right)^2}{2 m_\chi}
\eea
 
  More detailed flux estimation of the CRBDM for the secluded dark photon model and $U(1)_{B-L}$ model has been done thoroughly in~\cite{Guha:2024mjr} which includes the variation related to several values of DM mass and mediator mass, well spanned over the relevant parameter space. This analysis also agrees well with~\cite{Cao:2020bwd}.

\subsection{CRBDM Scattering at The Earth crust}
\label{sec:attenuation}

 Rigorous monte carlo simulation is beyond the scope of this work which helps to obtain the solution for the upper limit on the interaction cross section. The upper bound arises due to the Earth shielding on CRBDM while travelling towards the underground detector which attenuates the flux by virtue of the interaction with the various nucleus. We closely follow~\cite{Emken:2018run} and develop an feasible approach in~\cite{Guha:2024mjr}. At the depth of the detector ($d$), the kinetic energy of a CRBDM fermion are associated to the kinetic energy of the same at the surface of the Earth through the following well defined expression 
 \bea
 T_\chi^{\rm final}(d) = T_\chi^{\rm initial} \exp\Biggl[- \sum_N n_N \frac{2 \mu_{\chi N}^2 \sigma_{\chi N}}{m_N m_\chi} d - n_e \frac{2 \mu_{\chi e}^2 \sigma_{\chi e}}{m_e m_\chi} d \Biggr]\,,
 \label{eq:attenuated_KE}
 \eea
 where $n_N$ and $n_e$ are the number densities of nucleus species $N$ and electron in the Earth crust, respectively. 
 For the number densities of the nuclear species, we follow Ref.~\cite{Emken:2017qmp}~\footnote{$n_N = 2.7~\rm{g/cm^3} \cdot f_N \cdot N_A /A_N$ where $f_N, A_N$ are the mass fraction and the mass number of the species $N$ and $N_A$ is Avogadro's number.}. 
 On the other hand, we assume a constant number density of the electron, $n_e = 8 \times 10^{23}~\rm{cm^{-3}}$~\cite{Ema:2018bih}. 
 The DM-nucleus scattering cross section is related to the DM-nucleon scattering cross section via 
  \begin{align}
  \sigma_{\chi N} = \begin{cases}
     \sigma_{\chi n} A_N^2 \frac{\mu_{\chi N}^2}{\mu_{\chi n}^2} & \text{for the $U(1)_{B-L}$ model}\,, \\
   \mbox{\tiny\( \)}  & \mbox{\tiny\( \)} \\
      \sigma_{\chi n} Z_N^2 \frac{\mu_{\chi N}^2}{\mu_{\chi n}^2} & \text{for the secluded dark photon model}\,.
    \end{cases}
\label{Eq:nuclei_from_nucleon}
\end{align}
$A_N$ and $Z_N$ are the mass number and the atomic number of the nucleus $N$. $ \mu_{\chi N} = \frac{m_\chi m_N}{m_\chi + m_N}$ and $\mu_{\chi n} = \frac{m_\chi m_n}{m_\chi + m_n}$ are the reduced masses corresponding to the DM-nucleus and the DM-nucleon system.
 We estimate the attenuated CRBDM flux by the following expression~\cite{Bringmann:2018cvk}
 \bea
 \frac{d\Phi_\chi}{dT_\chi^{\rm final}}(d) = \frac{d\Phi_\chi}{dT_\chi^{\rm initial}}(d=0) \times \frac{dT_\chi^{\rm initial}}{dT_\chi^{\rm final}}\,.
 \eea
 Here $\frac{d\Phi_\chi}{dT_\chi^{\rm initial}}$ needs to be evaluated at $T_\chi^{\rm initial}$ where $T_\chi^{\rm initial}$ along with $\frac{dT_\chi^{\rm initial}}{dT_\chi^{\rm final}}$ can be evaluated using Eq.~(\ref{eq:attenuated_KE}) for any known $T_\chi^{\rm final}$.

In case of the energy-independent constant cross section scenario, we compute the attenuated CRBDM flux and we set the attenuation bound (upper limit on $\bar{\sigma}_{\chi}^{SI} $) as the value of $\bar{\sigma}_{\chi}^{SI} $ above which the CRBDM flux becomes negligible. But for the energy-dependent cross section scenario, this approach is not applicable. We modified the analysis and presented a concrete approximated version of the same in~\cite{Guha:2024mjr}. We use the similar approach in our present work as well. For different values of the coupling constant and the normalized cross section, we estimate the attenuated flux at the depth of the detector corresponding to the kinetic energy of CRBDM (with some particular $m_\chi$) which is the minimum value of the same to produce the detector threshold value of the nuclear recoil. At the last step, we find out the value of the coupling constant or the normalized cross section at which the attenuated CRBDM flux drops to a significantly lower value. Those values cause the CRBDM to have a higher number of interactions with the nuclei present inside the earth crust before reaching the detector, and are thus unable to produce nuclear recoil larger than the detector threshold. We set the upper bound by finding out such values at which the suppression starts.

\section{Results and Discussions}  
\label{sec:result}

Having equipped ourselves with the analytical description in the previous section, we now proceed with the numerical evaluation. We constrain the CRBDM parameter space and present the results in this section.

\subsection{Estimation of the differential event rate} 
\label{sec:event_rate}

Gathering all the bits and pieces, in the light of the discussion until now, we are ready to estimate the event rate at LZ. For CRBDM the differential recoil rate per target nucleus at the detector is given by~\cite{Bringmann:2018cvk}
\bea
\frac{dR}{dT_N} = \sigma^0_{\chi N} G_N^2(2 m_N T_N) \int^\infty_{T_\chi^{min}} \frac{dT_\chi}{T_N^{max}} \frac{d \Phi_\chi}{dT_\chi},
\eea
where, $\frac{d \Phi_\chi}{dT_\chi}$ be the CRBDM flux and $G_N^2(2 m_N T_N)$ be the nuclear form factor. Applicability of the above expression is validated only for constant cross section consideration. For specific underlying model the following modification has to be done
\bea
\frac{dR}{dT_N} = G_N^2(2 m_N T_N) \int^\infty_{T_\chi^{min}} dT_\chi \frac{d\sigma_{\chi N}}{dT_N} \frac{d \Phi_\chi}{dT_\chi}.
\eea
In that case the total number of events per target nucleus per unit time in CRBDM scenario is given by
\bea
R = \int_{T_1}^{T_2} dT_N G_N^2(2 m_N T_N) \int^\infty_{T_\chi^{min}} dT_\chi \frac{d\sigma_{\chi N}}{dT_N} \frac{d \Phi_\chi}{dT_\chi};
\label{eq:rate_ener_dependent}
\eea
with, $\left\lbrace T_1, T_2 \right\rbrace $ be the accessible window of the nuclear recoil energy. For LZ experiment, efficiency weighted $\left\lbrace T_1, T_2 \right\rbrace = \left\lbrace 2, 70 \right\rbrace $ keV ~\cite{LZ:2022lsv}. No additional events over the background are observed at LZ experiment with the fiducial mass of 5.5 tonnes and the exposure livetime for 60 days. For null observation, the number of events at the $90\%$ CL corresponds to 2.3. Now utilizing the CRBDM flux, we obtain the $90\%$ CL limit by equating the expected total number of CRBDM events to the efficiency function weighted event rate integration, i.e., the predicted total number of events by the model.
  
  With the information stated above the actual number of predicted event turns out to be
\bea
N_R(g_i,g_\chi, m_\chi, m_{A^{\prime}}) = N_t \times exposureLZ \int_{T_1}^{T_2} dT_N\, eff(T_N) G_N^2(2 m_N T_N) \int^\infty_{T_\chi^{min}} dT_\chi \frac{d\sigma_{\chi N}}{dT_N} \frac{d \Phi_\chi}{dT_\chi},
 \label{Eq:diff_rate}
\eea
where, $eff(T_N)$ is the efficiency of the LZ experiment, $N_t= 5.5~\rm{tonnes}/m_{Xe}$(mass of Xenon nucleus) is the number of target particles and $exposureLZ = 60 \times 24 \times 3600~\rm{s}$.
With the overall coupling $g_i g_\chi$, the differential cross section is defined as
\bea
\frac{d\sigma_{\chi N}}{dT_N} = \frac{g_i^2 g_\chi^2}{4 \pi}~\frac{2 m_N \left(m_\chi + T_\chi \right)^2 - T_N \left \lbrace \left(m_\chi + m_N \right)^2 + 2 m_N T_\chi \right\rbrace + m_N T_N^2}{\left(2 m_N T_N + m_{A^{\prime}}^2 \right)^2  \left(2 m_\chi T_\chi + T_\chi^2 \right)}.
\eea

\subsection{Constraining the coupling}  
\label{sec:constraint_eps}
 
 We constrain the overall coupling for the two models mentioned above, using the LZ data~\cite{LZ:2022lsv}. For our analysis we assume $g_\chi = 1$ (the coupling between the DM and the mediator). We estimate the lower bound on the coupling between the SM nucleon and the mediator($g_i$) as follows : 
\bea
g_i  &=& \left(\frac{2.3}{N_R(g_i=1,g_\chi = 1, m_\chi, m_{A^{\prime}}) } \right)^{1/4}. 
\eea
We also explore an alternative approach in order to constrain the couplings. Following~\cite{Bringmann:2018cvk} we write the total event rate per target nucleus for conventional DD experiments (for cold DM) as follows
\bea
R = \kappa \frac{\sigma^{CDM}_{\chi N}}{m_{CDM}} \left(\bar{v} \rho_{CDM} \right)^{local} ~~~ \rm{for} ~~ m_{CDM} \gg m_N ,
\label{Eq:total-R-CDM}
\eea
with
\bea
\kappa = \exp[-2T_1/(\pi m_N \bar{v}^2)] - \exp[-2T_2/(\pi m_N \bar{v}^2)].
\eea
The most stringent limit for the spin-independent scattering has been reported by LZ experiment at $m_{CDM}=$ 36 GeV which rejects cross sections above $9.2 \times 10^{-48}~\rm{cm^2}$ at the $90\%$ confidence level. Using these values in Eq.~(\ref{Eq:total-R-CDM}) and equating that to the expression at Eq.~(\ref{eq:rate_ener_dependent}) we obtain the limiting value of the scattering rate per nucleon (for constant cross section scenario)
\bea
\sigma^{BDM.limit}_{\chi} &=& \sqrt{\kappa \left(\bar{v} \rho_{CDM} \right)^{local} \left( \frac{m_\chi + m_N}{m_\chi + m_p}\right)^2 \left(\frac{\sigma^{CDM}_{\chi nucleon}}{m_{CDM}} \right)_{m_{CDM} \gg m_N} }\nonumber \\
&\times & \left(\int_{T_1}^{T_2} dT_N \int^\infty_{T_\chi^{min}} \frac{dT_\chi}{T_N^{max}} \left\lvert\frac{d \Phi_\chi}{dT_\chi}\right\rvert_{\sigma^{BDM}_{\chi}=1} \right)^{-1/2}.
\eea
For specific model this turns out to be
\bea
g_{B-L} ~\rm{or}~ e \varepsilon \cos \theta_W  &=& \left[{\kappa \left(\bar{v} \rho_{CDM} \right)^{local} \left( \frac{m_\chi + m_N}{m_\chi + m_p}\right)^2 \left(\frac{\sigma^{CDM}_{\chi nucleon}}{m_{CDM}} \right)_{m_{CDM} \gg m_N}} \right]^{1/4} \nonumber \\
&\times & \left(\int_{T_1}^{T_2} dT_N \int^\infty_{T_\chi^{min}} dT_\chi \left\lvert \frac{d\sigma_{\chi N}}{dT_N}\right\rvert_{g_{B-L} ~\rm{or}~ e \varepsilon \cos \theta_W = 1} \left\lvert \frac{d \Phi_\chi}{dT_\chi} \right\rvert_{g_{B-L} ~\rm{or}~ e \varepsilon \cos \theta_W = 1} \right)^{-1/4} \nonumber \\
\eea

\begin{figure*}
 \centering
 \includegraphics[width=0.49 \textwidth]{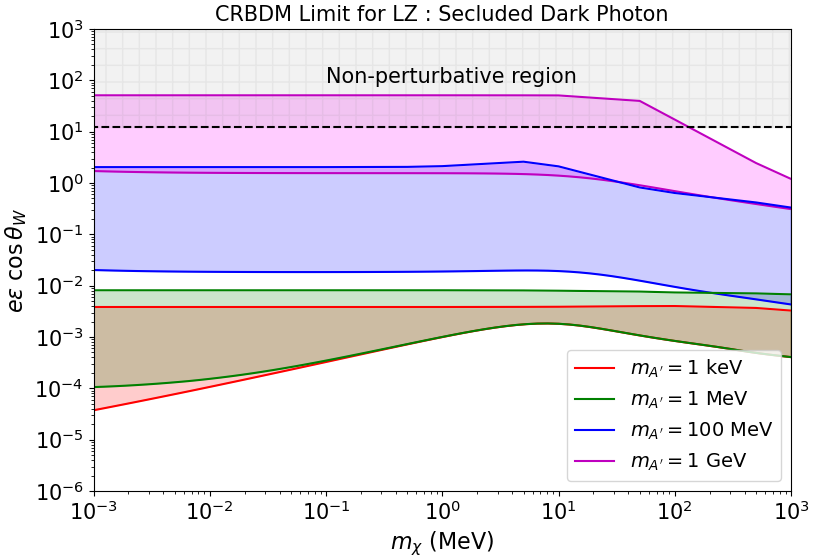}
 \includegraphics[width=0.49 \textwidth]{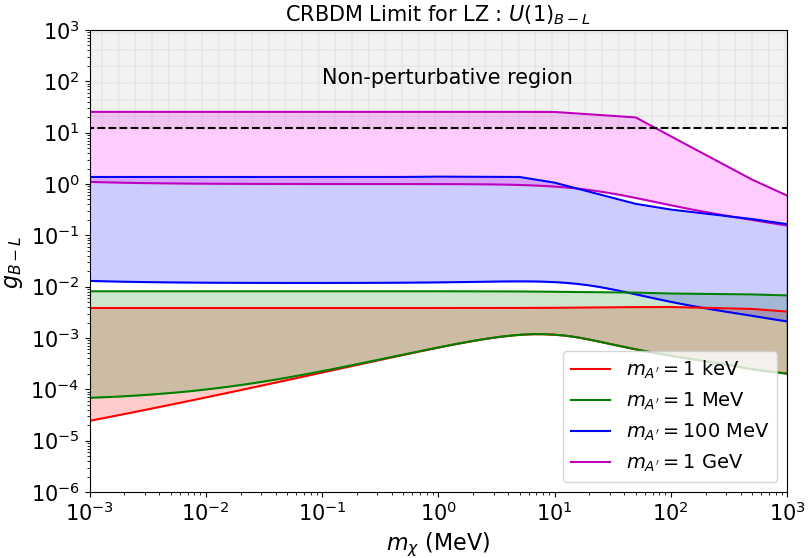}
 \caption{Exclusion limits at $90\%$ confidence level for $g_\chi = 1$, in the coupling vs $m_\chi$ plane for secluded dark photon model (left) and $U(1)_{B-L}$ model (right) with different values of the mediator mass. Dotted lines with the same color code stands for the upper limit due to the earth shielding effect.
 }
 \label{fig:coup_constraint_LMHM}
 \end{figure*}

\begin{figure*}
 \centering
 \includegraphics[width=0.49 \textwidth]{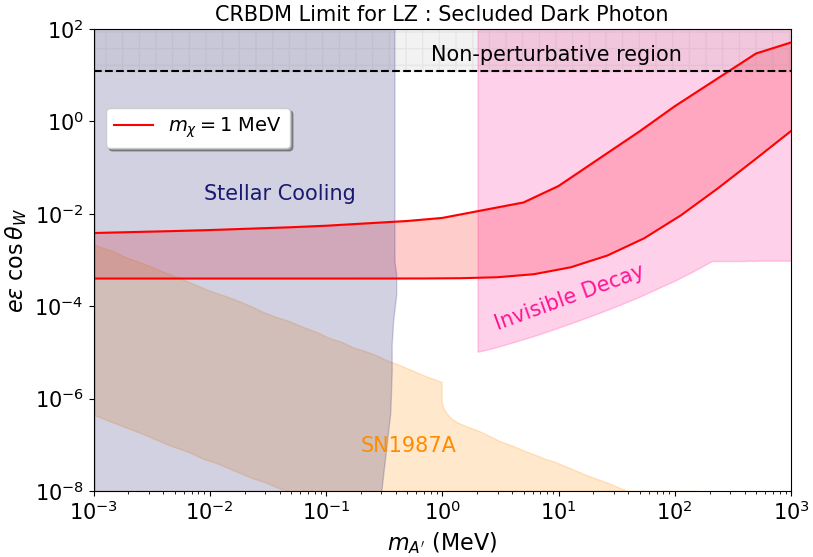}
 \includegraphics[width=0.49 \textwidth]{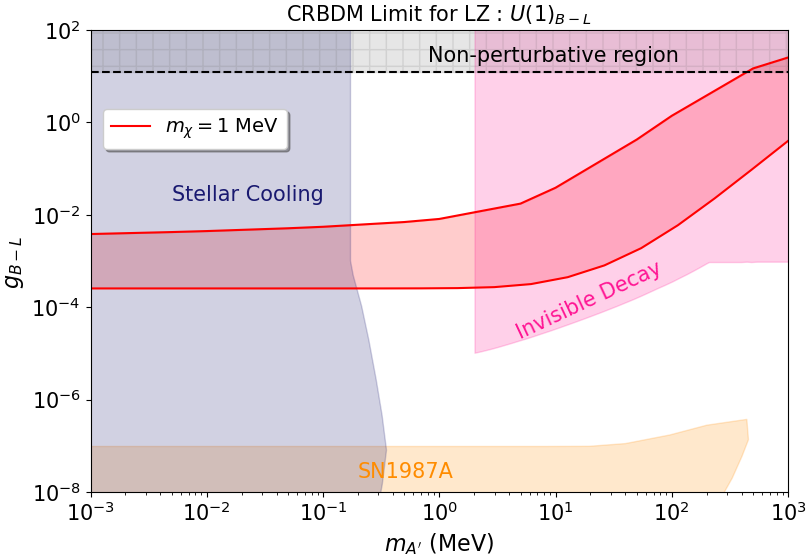}\\
 \includegraphics[width=0.49 \textwidth]{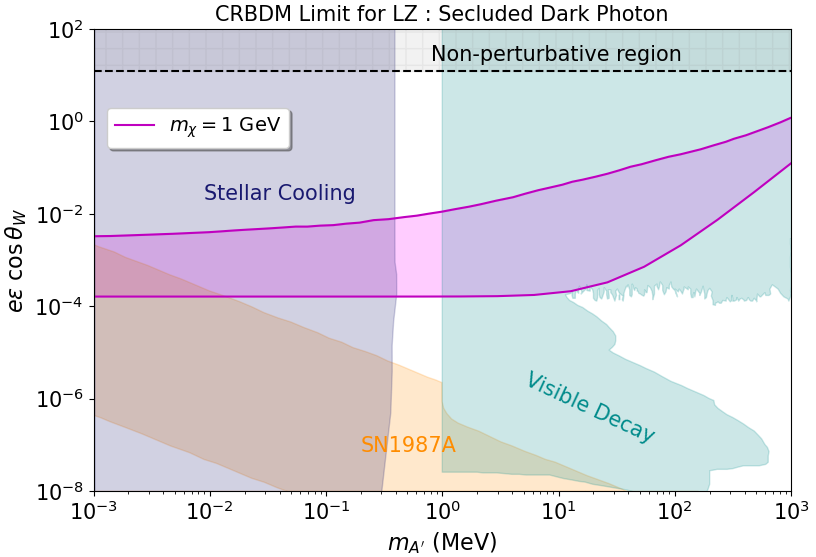}
 \includegraphics[width=0.49 \textwidth]{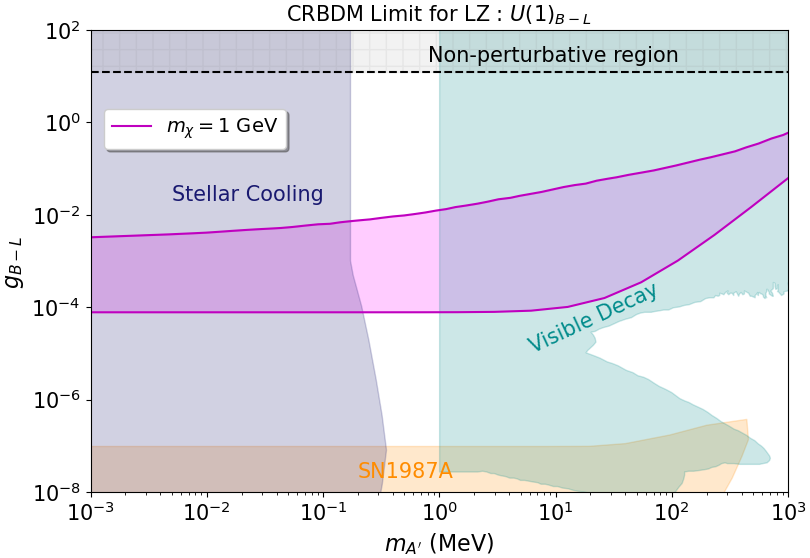} 
 \caption{Exclusion limits at $90\%$ confidence level for $g_\chi = 1$, in the coupling vs $m_{A^{\prime}}$ plane for secluded dark photon model (left) and $U(1)_{B-L}$ model (right) with different values of the DM mass. Dotted lines with the same color code stands for the upper limit due to the earth shielding effect. In the upper and lower panel we show our results along with the existing limits. Invisible decay denotes the scenario where the mediator dominantly decays to the dark sector whereas the dominant decay to the SM sector is defined by the visible decay.
 }
 \label{fig:coup_constraint_mvecA}
 \end{figure*}

In general this expression is dependent on the mass of the mediator. We found that the two different approaches to constrain the overall coupling (described above) yield quite identical results. In order to avoid repetitions we stick to the first approach and report the obtained lower limit of the exclusion region along with the upper limit of the same due to the Earth shielding effect in Fig.~\ref{fig:coup_constraint_LMHM} and in Fig.~\ref{fig:coup_constraint_mvecA}. Specifically we show the variation of the exclusion region with respect to the DM mass keeping the mass of the mediator fixed to certain values in Fig.~\ref{fig:coup_constraint_LMHM}.

On the other hand, we fix the DM mass to some constant values and show the variation of the exclusion region with respect to the mediator mass in Fig.~\ref{fig:coup_constraint_mvecA}. This part of the analysis agrees well with~\cite{Bell:2023sdq}.
We also show the existing limits for comparison~\cite{Bauer:2018onh}. 
In Ref.~\cite{Bauer:2018onh}, the visible decay bounds have small variations for two different models. 
For the invisible decay case, we show the existing exclusion limit in pink. 
The bounds on the invisible decay of the hidden photon are from NA64~\cite{Andreev:2021fzd,Banerjee:2019pds}, NA62~\cite{NA62:2019meo}, Babar~\cite{Filippi:2019lfq,BaBar:2017tiz}, E787, and E949~\cite{Essig:2013vha}. 
The exclusion limits on the decay of dark photons into the pair of electron-positron ($e^+ e^-$) are from Babar~\cite{Filippi:2019lfq,BaBar:2014zli}, NA64~\cite{NA64:2019auh}, NA48~\cite{NA482:2015wmo}, E141~\cite{Fabbrichesi:2020wbt,Riordan:1987aw}, E137~\cite{Fabbrichesi:2020wbt,Bjorken:1988as,Batell:2014mga,Marsicano:2018krp}, E774~\cite{Bross:1989mp}, nu-Cal~\cite{Blumlein:2011mv,Blumlein:2013cua}, CHARM~\cite{Gninenko:2012eq}, KLOE~\cite{KLOE-2:2012lii,KLOE-2:2014qxg,Anastasi:2015qla}, WASA~\cite{WASA-at-COSY:2013zom}, HADES~\cite{HADES:2013nab}, APEX~\cite{APEX:2011dww}, A1~\cite{Merkel:2014avp},BESII~\cite{BESIII:2017fwv}, and LSND~\cite{LSND:1997vqj}. Bounds obtained from the Meson decay~\cite{Dror:2017ehi} (considering the further visible decay of new boson) are overlapping with the above mentioned bounds (not shown in the figure). The stellar cooling bound and the SN1987A cooling bound for the secluded dark photon model are obtained from \cite{Chang:2016ntp}. Whereas, the same bounds for the $U(1)_{B-L}$ model are obtained from \cite{Hardy:2016kme} and \cite{Rrapaj:2015wgs}, respectively. 
 
\subsection{Constraining the cross section with the knowledge of the exclusion limit on the coupling} 
\label{sec:cs_constraint}

In order to convert the limit on the coupling to the limit on the interaction cross section, we normalize the DM-nucleon scattering cross section ${\sigma}_{\chi}^{SI}$ with the following definitions~\cite{Cao:2020bwd,Essig:2011nj}
\bea
\overline{\abs{\mathcal{M}_{free}}^2} &=& \overline{\abs{\mathcal{M}_{free}(q_{ref})}^2} \times \abs{F_{DM}(q)}^2 \nonumber \\
{\sigma}_{\chi}^{SI} &=& \frac{\mu_{\chi n}^2~\overline{\abs{\mathcal{M}_{free}(q_{ref})}^2}}{16 \pi m_\chi^2 m_n^2} 
\eea
Where, the form factor is given by (at the detector)
%
\bea
\abs{F_{DM}(q)}^2 = \frac{(q_{ref}^2 + m_{A'}^2)^2}{(2 m_n E_R + m_{A'}^2)^2} \times \frac{2 m_n (m_\chi + T_\chi)^2 - E_R \left[ (m_\chi + m_n)^2+ 2 m_n T_\chi\right] + m_n E_R^2}{2 m_n m_\chi^2} \nonumber \\
\eea 
%
In the non-relativistic limit $T_N, T_\chi \ll m_n$
\bea
\abs{F_{DM}(q)}^2 = \frac{(q_{ref}^2 + m_{A'}^2)^2}{(q^2 + m_{A'}^2)^2} 
\eea 
For heavy-mediator limit $\abs{F_{DM}(q)} = 1$ and for light-mediator limit $\abs{F_{DM}(q)} \sim \frac{q_{ref}^2}{q^2}$. With these definitions
%
\bea
\overline{\abs{\mathcal{M}_{free}(q_{ref})}^2} &=& \frac{16 {g_{n}^{A'}}^2  {g_{\chi}^{A'}}^2 m_n^2 m_\chi^2}{(q_{ref}^2 + m_{A'}^2)^2} \nonumber \\
{\sigma}_{\chi}^{SI}  &=& \frac{{g_{n}^{A'}}^2{g_{\chi}^{A'}}^2 \mu_{\chi n}^2}{\pi (q_{ref}^2  + m_{A'}^2)^2} \nonumber \\
\implies {\sigma}_{\chi}^{SI}  &=& \begin{cases}
        \frac{{g_{n}^{A'}}^2{g_{\chi}^{A'}}^2 \mu_{\chi n}^2}{\pi (q_{ref}^2 )^2} ~~~\text{for light mediator,}
        \\
        \frac{{g_{n}^{A'}}^2{g_{\chi}^{A'}}^2 \mu_{\chi n}^2}{\pi (m_{A'}^2)^2} ~~~\text{for heavy mediator}.
        \end{cases}
\eea
%
The definition used to covert the bound on couplings to the bound on cross section is as follows (valid only for heavy mediator scenario):
\bea
{\sigma}_{\chi}^{SI}  = \frac{\mu_{\chi p}^2 g_i^2 g_\chi^2}{\pi m_{A^{\prime}}^4}.
\eea

 \begin{figure*}
 \centering
 \includegraphics[width=0.49 \textwidth]{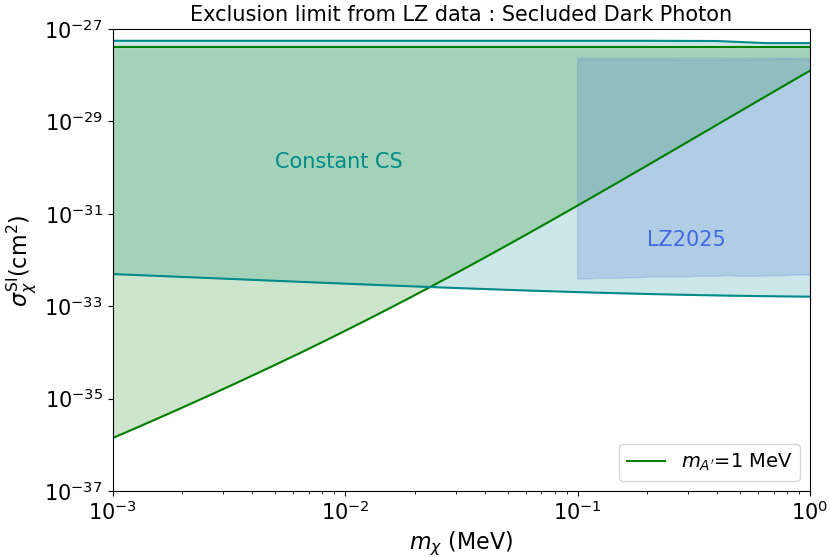}
 \includegraphics[width=0.49 \textwidth]{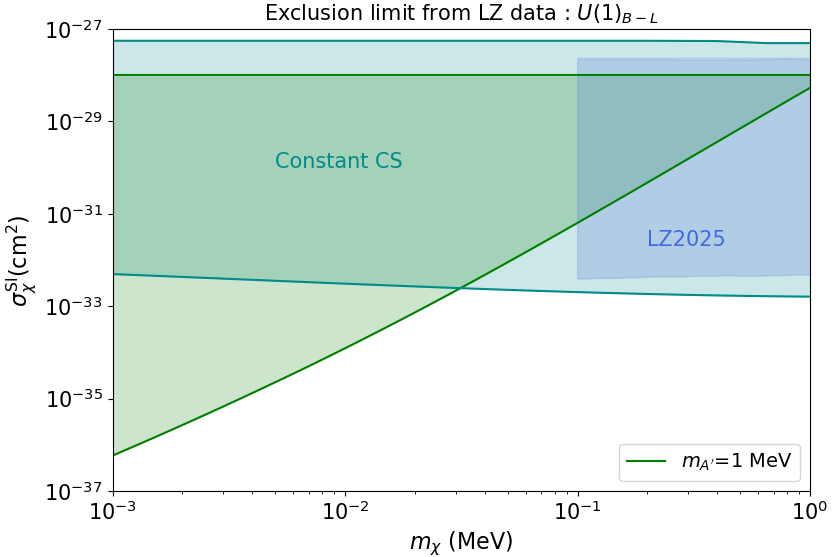}\\
 \includegraphics[width=0.49 \textwidth]{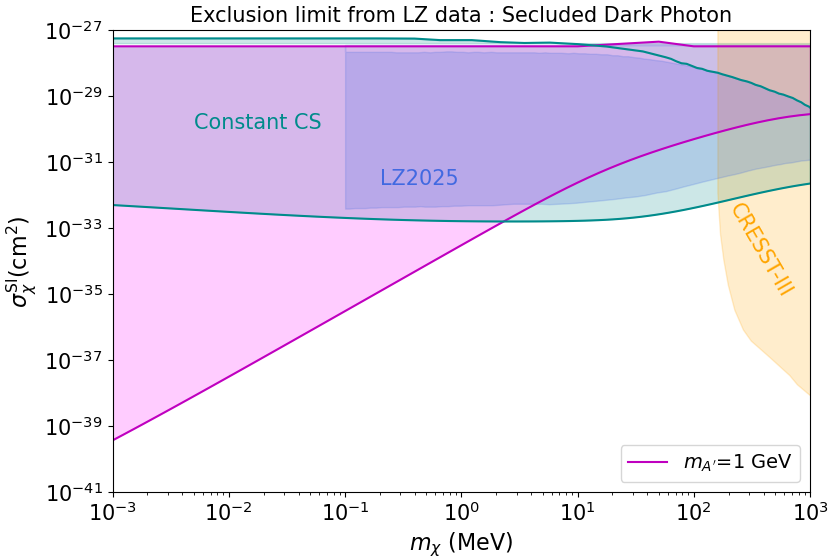}
 \includegraphics[width=0.49 \textwidth]{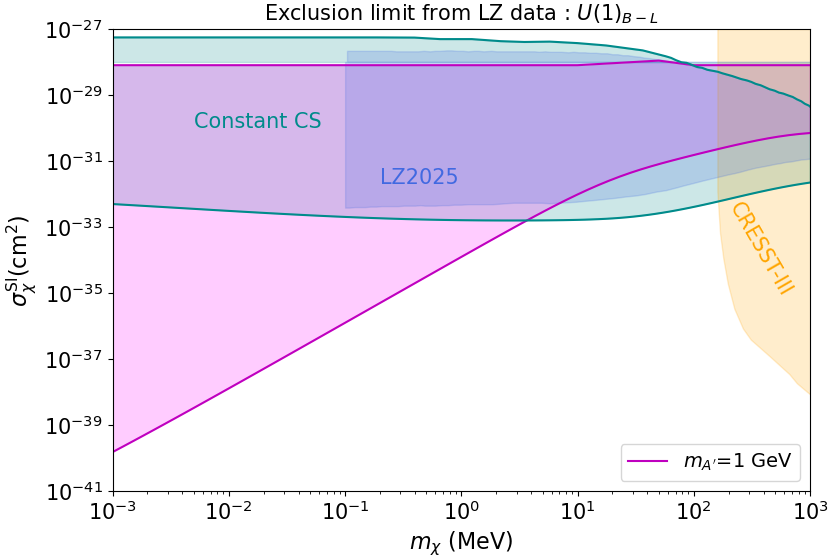} 
 \caption{Exclusion limits at $90\%$ confidence level in the ${\sigma}_{\chi}^{SI}$ vs $m_\chi$ plane for secluded dark photon model (left) and $U(1)_{B-L}$ model (right) with different values of the mediator mass. The constant cross section scenario is defined by the consideration of energy independent cross section without any underlying model, $\frac{d\sigma_{\chi n}}{dT_\chi} = \frac{\sigma_{\chi n}}{T_\chi^{max}}$ and $\frac{d\sigma_{\chi n}}{dT_N} = \frac{\sigma_{\chi n}}{T_N^{max}}$. For comparison we also show the bounds obtained from CRESST-III (conventional DD of cold DM)~\cite{CRESST:2019jnq} and LUX-ZEPLIN (CRBDM)~\cite{LZ:2025iaw} along with our estimated limits.
}
 \label{fig:sigmae_constraint_LMHM}
 \end{figure*}

We show the variation of the exclusion region of the cross section along with the mass of the DM for different values of the mass of the mediator in Fig.~\ref{fig:sigmae_constraint_LMHM}. For comparison we also show the bound obtained with the consideration of energy independent cross section without any specific underlying model which is denoted as the constant cross section scenario. Our limits in the constant cross section scenario agree with the recent bound reported by LUX-ZEPLIN considering the CRBDM formalism~\cite{LZ:2025iaw}. The exclusion region reported by PANDAX-II~\cite{PandaX-II:2021kai}, considering similar CRBDM formalism, falls within the LUX-ZEPLIN exclusion limit. For a more realistic description, we extended our work to the energy dependent cross section scenario considering the models described before. We also include the existing conventional direct detection limit as a yellow shaded region, which is taken from CRESST-III~\cite{CRESST:2019jnq}.

\section{Conclusion}
\label{sec:conclusion}

We obtained the exclusion limit on the DM-nucleon spin-independent cross section employing the LZ data and considering CRBDM mechanism. We found that this approach helps us to explore much lower mass region of DM parameter space compared to conventional DM search strategies. In the process of estimating the same, we work on the framework of two well-known models containing new $U(1)'$ symmetry, namely, secluded dark photon model and $U(1)_{B-L}$ model. The upper bound of the exclusion region is comparable in both the energy dependent and energy independent cases. On the other hand, there is significant improvement in case of the lower bound while considering the specific energy dependence invoked by the underlying physics model, as compared to the energy independent constant cross section scenario. For example, if we consider $m_\chi =$ 1 keV, the constant cross section consideration yields the lower bound on $\sigma_\chi^{\rm{SI}} \approx 4\times 10^{-33}~\rm{cm^2}$, while the secluded dark photon model and the $U(1)_{B-L}$ model give the lower bound on $\sigma_\chi^{\rm{SI}}$ to be $\approx 6.4\times 10^{-40}~\rm{cm^2}$ and $2.5\times 10^{-40}~\rm{cm^2}$, respectively, for $m_{A^{\prime}}$=1 GeV. The outcome of the energy independent cross section scenario agrees well with the recent CRBDM constraints reported by LUX-ZEPLIN. According to these newest constraints, the exclusion region of the spin-independent CRBDM-nucleon cross section is bounded by the value $3.9\times 10^{-33}~\rm{cm^2}$ on the lower side for sub-GeV masses of DM~\cite{LZ:2025iaw}. Whereas most of the conventional DD experiments lose their sensitivity at DM mass of 1 GeV, we could explore the lower mass range of DM, i.e., 1 keV-1 GeV mass range considering the CRBDM framework. Although the constant cross section consideration gives us quick estimates of the limiting values of the exclusion region but for more realistic scenario one must work on well motivated models which are helpful to obtain much realistic and trustworthy bounds.

\acknowledgments 
This work was supported by the National Research Foundation of Korea grant funded by the Korea government (MSIT) (RS-2024-00356960).

\bibliographystyle{JHEP}
\bibliography{ref}

\end{document}